\documentclass[prl,superscriptaddress,twocolumn,amsmath,amssymb]{revtex4-1}
\usepackage{graphicx}
\usepackage{float}
\usepackage{color}
\usepackage{epstopdf,cancel,ulem}
\usepackage{epsf,latexsym,bbm,euscript}
\usepackage{amssymb,amsmath}
\usepackage{mathtools} %used in this file for the command \coloneqq -> :=
\usepackage{times,graphics}
\usepackage{soul,xcolor}
\usepackage{epsfig}

\newcommand{\be}{\begin{equation}}
\newcommand{\ee}{\end{equation}}
\newcommand{\ba}{\begin{eqnarray}}
\newcommand{\ea}{\end{eqnarray}}

\newcommand{\beq}{\begin{equation}}
\newcommand{\eeq}{\end{equation}}
\newcommand{\beqa}{\begin{eqnarray}}
\newcommand{\eeqa}{\end{eqnarray}}

\def\6{\langle}
\def\9{\rangle}

\def\be{\begin{equation}}
\def\ee{\end{equation}}

\def\bali{\begin{align}}
\def\eni{{\end{align}}}

\def\1{{{\mathbbm 1}}}

\def\pad{{\partial}}

\def\sg{\textsl{g}}

\begin{document}
\title{ Horizon avoidance in spherically-symmetric collapse}

\author{Valentina Baccetti}
\affiliation{Department of Physics \& Astronomy, Macquarie University, Sydney NSW 2109, Australia}
\author{Robert B. Mann}
\affiliation{Department of Physics and Astronomy, University of Waterloo, Waterloo, Ontario, Canada}
\affiliation{Perimeter Institute for Theoretical Physics, Waterloo, Ontario, Canada}
\author{Daniel R. Terno}
\affiliation{Department of Physics \& Astronomy, Macquarie University, Sydney NSW 2109, Australia}

%\email{daniel.terno@mq.edu.au}
\begin{abstract}
 We study collapse of evaporating spherically-symmetric thin dust shells and dust balls assuming that quantum effects are encapsulated in a spherically-symmetric metric that satisfied mild regularity conditions.  The evaporation may accelerate  collapse, but for a generic metric the Schwarzschild radius  is not crossed. Instead the shell (or the layer in the ball of dust) is always at a certain sub-Planckian distance from it.
\end{abstract}
\maketitle

\textit{Introduction.}---
Event horizon --- the null surface that bounds the spacetime region from which signals cannot escape  --- is the defining feature of black holes in general relativity \cite{fn:98,poisson, ast-bh}. This classical concept plays an important role in their quantum {behaviour} \cite{fn:98,h:74,wald:01, bmps:95,visser:08,kiefer:07,modern}. The emission of Hawking radiation completes a thermodynamic picture of black holes, and its most straightforward derivation relies on the existence of a horizon. Radiation also leads to the black hole information loss paradox \cite{h:76}, perhaps  the longest-running controversy in theoretical physics  \cite{kiefer:07,modern,h:76}.

{The event horizon underpins the paradox, which can be stated as follows \cite{wald:01}. The initial state of the collapsing
matter has a low non-zero entropy \cite{ht:10,bht:17} and the black hole it  forms subsequently evaporates  within a finite time.
This evolution of  low-entropy matter into  high-entropy radiation implies that some  information will have been lost unless the correlations between the inside and outside of  the horizon can be restored during this process.}

Multiple proposals exist to recover the information and exorcise or dismiss the paradox, leading to   numerous controversies and counterclaims (see for example  \cite{kiefer:07,modern, bht:17}). Here we focus instead on  the event horizon, whose existence may not survive  quantum effects \cite{brust:14}.

We first recall    that in   general relativity    collapse of a classical matter distribution into a black hole, or infall of a test particle into it, takes an infinite amount of time according to the clock of  Bob  (an observer at  spatial infinity) \cite{fn:98,poisson}. Thus formation of the event horizon or its crossing are in principle unobservable by external observers \cite{visser:14}.   However, the same  collapsing matter or a test particle crosses the horizon in a finite proper time. This not only proves that the coordinate systems that use the time at infinity are geodesically incomplete, but establishes   the event horizon as a physical object and not just  a useful mathematical concept.

Next we note that while the event horizon is an essential ingredient for obtaining Hawking radiation in a static case \cite{bmps:95}, it is not necessary in the dynamical spacetime of a collapsing distribution of matter \cite{visser:08,haj:86}.  Both numerical studies of   collapsing shells and  analytic  results  \cite{haj:86,acvv:01,blsv:06,vsk:07,kmy:13,ss:15}  indicate  the existence of a pre-Hawking radiation.

Finally,  analyses of the effects of radiation on thin null  \cite{kmy:13} and   massive \cite{bmt:16} shells  demonstrate that taking this pre-Hawking radiation into account  may  {preclude the existence of an event horizon. For example collapsing massive shells, which allow the  introduction of a co-moving observer (Alice), were shown in two specific spherically-symmetric models to each have  receding Schwarzschild radii that cannot be crossed in Alice's  finite proper time (or in other words in time measured by any clock co-moving with the collapsing shell) \cite{bht:17,bmt:16}. The shell thus evaporates before the horizon can ever actually form.}

{How generic is such horizon avoidance? This natural question is particularly} important since there is no ready-made prescription that provides  the expectation value of the stress-energy tensor that  feeds into the Einstein equations and self-consistently produces a metric.
Here we investigate this question and show that for a generic  spherically-symmetric metric  that results from an early onset of  radiation the formation of an event horizon is avoided.

We first present the classical setting for massive thin shell collapse, and then review the derivation for the case where the outside spacetime geometry can be modelled by the outgoing Vaidya metric. It serves as a template for the subsequent analysis of a general spherically-symmetric metric. Finally we extend these results to the collapse of a   ball of dust.   We use $(-+++)$ signature of the metric and set $c=\hbar=G=k_B=1$.

\textit{The classical model}--- We  briefly review   the relevant aspects of the  collapse  of a  massive  spherically-symmetric thin  shell $\Sigma$
  in $3+1$ dimensional spacetime \cite{poisson}.  All the results are straightforwardly generalized to any $D>3$ spatial dimension.

The spacetime inside the shell is flat \cite{poisson}. In  the absence of Hawking radiation  the metric
 \begin{eqnarray}
 ds^2_+ &=&- f(r_+) dt^2_+ + f(r_+)^{-1}dr^2_++r^{2}_+ d\Omega  \label{met0}\\
 &=& -f(r_+) du_+^2-2du_+ dr_+ + r_+^{2} {d\Omega},
  \label{met2}
 \end{eqnarray}
describes the exterior geometry in terms of standard coordinates and  outgoing Eddington-Finkelstein coordinates $(u_+,r_+)$, respectively. Here  $dt _+ = du_+ +dr_+/f(r_+)$,
 $f(r)=1-C/r$ (in three spatial dimensions $C=2M=r_\sg$ is the Schwarzschild radius),  {and $d\Omega$ is the spherical surface element}. Their counterparts in the Minkowski spacetime  inside the shell are $(t_-,r_-)$ and $(u_-,r_-)$, where $u_-=t_- - r_-$.

%The angular coordinates $(\phi_1,\ldots,\phi_{D-1})$ are discussed in Supplementary Material, Sec.~I.

The shell's trajectory is parameterized by its proper time $\tau$  as, e. g.,    $\big(U_\pm(\tau), R_\pm(\tau)\big)$ using  the $(u,r)$ coordinates outside and inside the shell.

The first junction condition \cite{poisson, isr:66}  is the statement that the induced metric $h_{ab}$ on the shell is the same on  both sides $\Sigma^\pm$, $ds^2_\Sigma=h_{ab}dy^ady^b=-d\tau^2+R^2d\Omega$. It leads to the identification $R_+(\tau)\equiv R_-(\tau)=:R(\tau)$. Since for massive particles the four-velocity $v^\mu$ satisfies $v_\mu v^\mu=-1$, shell's trajectory obeys
\be
\dot U_+=\frac{-\dot R+\sqrt{F+\dot R^2}}{F}, \label{udot}
\ee
where   $\dot A=dA/d\tau$ and $F:=1-C/R$.

Discontinuity of the extrinsic curvature $K_{ab}$ is described by the second junction condition \cite{poisson,isr:66}
\be
S_{ab}=-\big([K_{ab}]-[K]h_{ab}\big)/8\pi, \label{eqofmot}
\ee
where {$K:=K^a_{\,a}$, and }$[A]:=A|_{\Sigma^+}-A|_{\Sigma^-}$ is the discontinuity of the quantity $A$ across the two sides $\Sigma^\pm$ of the surface. %The surface stress-energy tensor for a thin  dust shell   is
%\be
%S^{ab}=\sigma v^a v^b=\sigma \delta^a_\tau\delta^b_\tau,
%\ee
%where $\sigma$ is the surface density and $v^a$ are the components of the proper velocity  in the surface coordinates $y^a$.
The resulting equation of motion is  particularly simple for dust (an ideal pressureless fluid),
\begin{eqnarray}
  {\mathcal{D}(R) } &:=&  \frac{2\ddot R + F'}{2\sqrt{F+\dot R^2}} - \frac{\ddot R}{\sqrt{1+\dot R^2}} \nonumber\\
&& + (D-2)\frac{\sqrt{F+ \dot R^2} - \sqrt{1+\dot R^2}}{R} = 0 \label{sangC}
\end{eqnarray}
where   $A'=\pad A/\pad r |_\Sigma$. This equation is  simple enough to have an analytic solution $\tau(R)$, {leading to the finite crossing time $\tau(r_\sg)$.}

\textit{Post-evaporation metric: a special example} --- The spacetime inside the shell is still Minkowskian, while we assume that evaporation leads to   the outgoing Vaidya metric \cite{vai:51} outside it.  It  is  {given by \eqref{met2}
but with  $f\to f(u,r)=1-C(u)/r$, where we only assume that $C\geq 0$ and $dC/du\leq0$}.
Here and in the following we drop the subscript ``+" . This metric is often used to describe the exterior of evaporating black holes \cite{bmps:95,kmy:13,fw:99}.

This  example   provides us with the template for a general analysis.  By monitoring the gap between the shell and the Schwarzschild radius,
\be
x(\tau):=R(\tau)-r_\sg\big(U(\tau)\big),
\ee
we discover  how evaporation modifies the classical shell dynamics.

The equation of motion of the shell  {\eqref{sangC} becomes} \cite{bmt:16}
\begin{eqnarray}
{\mathcal{D}(R)} -F_U \dot{ U} \left( \frac{\dot R}{2F\sqrt{F+\dot R^2}} -\frac{1}{2F}\right)= 0,
\label{eq:eom-vaidya}
\end{eqnarray}
where
$A_u:=dA/du$, and $\dot U$ is given by Eq.~\eqref{udot}. Solving Eq.~\eqref{eq:eom-vaidya} for $\ddot R$, then expanding in   inverse powers of $x$ and $C$
near the Schwarzschild radius we find that the collapse accelerates,
\be
\ddot R \approx \frac{4 \dot R^4 C}{   x^2}\frac{d C}{d U} < 0.
\ee
while the rate of approach to the Schwarzschild radius, $\dot x$, behaves as
\be
\dot x\approx \dot R\big(1-\epsilon_*(\tau)/x(\tau)\big), \label{xeps}
\ee
where a natural time-dependent scale for this problem is
\be
\epsilon_*:= {2C} \left|\frac{d C}{d U}\right|.
\ee

The gap decreases only as long as $\epsilon_*<x$. If the evaporation law is such that this is true for its entire duration, we have $R>r_g$ until the evaporation is complete. Otherwise, once the distance between the shell and the Schwarzschild radius is reduced to $\epsilon_*$, it cannot decrease any further.

From the point of view of Bob the shell is stuck within a slowly changing coordinate distance $\epsilon_*$ from the slowly receding Schwarzschild radius. For Alice the collapse accelerates,  {but never leads to horizon crossing and thus never to horizon formation.}  Only if the evaporation rate $dC/du$  vanishes the equation of motion Eq.~\eqref{eq:eom-vaidya} reduces to Eq.~\eqref{sangC} as $\epsilon_*\rightarrow 0$ \cite{bmt:16}.
%%%%%%%%%%%%%%%%

\textit{General spherically-symmetric metric.}--- We do not assume any particular  form of the expectation of the stress-energy tensor $\6\hat{T}_{\mu\nu}\9$ beyond   spherical symmetry.  The geometry outside the shell is therefore described by a general spherically-symmetric metric. In  $(u,r)$ coordinates it is given by \cite{blau}
\be
ds^2=-e^{2h(u,r)}f(u,r)du^2-2e^{ h(u,r)}dudr+r^2d\Omega \label{met-gen}
\ee
where the two  functions $f(u,r)$ and $h(u,r)$ should satisfy certain mild restrictions.
It is convenient to introduce the mass function $C(u,r)$ as
\be
f=:1-{C(u,r)}/r
\ee
The outgoing Vaidya metric corresponds to $C(u,r)=C(u)=r_\sg(u)$ and $h(u,r)=0$.

We assume that (i) $0\leq C<\infty$ with $C(u,r)>0$  for $u<u_E<\infty$, and   $\pad C/\pad  u<0$  as long as the function $C>0$   to ensure a  finite positive gravitational mass, positive energy density and   positive flux at infinity; (ii) $h(u,r)$ is continuous; (iii) the metric has only one coordinate singularity, namely an (infinite red-shift)   surface $f(u,r)=0$. This surface is located at the Schwarzschild radius   $r_\sg(u)$, that  is   implicitly given by
\be
r_\sg\equiv C(u,r_\sg). \label{rgdef}
\ee
%We also assume that (iv) $f(u,r)>0$  for $r>r_\sg$.

From (iii) it follows that
\be
C(u,r)=r_\sg(u)+ w(u,r)\big(r-r_\sg(u)\big), \label{kdecomp}
\ee
for some function $w(u,r)$. Assumption (i) implies that  $w(u,r_\sg)\leq 1$.

For large $r$ assumption (i) leads to $f\rightarrow 1$, and since by (ii) it has the same sign for $r>r_\sg$, we find that everywhere outside the Schwarzschild radius
\be
f(u,r)>0,  \qquad r>r_\sg. \label{posf}
\ee
This is the same as $C(u,r)<r$, and thus $w(u,r)<1$ for $r>r_\sg$.
%\tcg{[\textsc{Comment: the bound $C<\infty$ in (i) allows to avoid postulating (iv) $f>0$ for $r>r_\sg$}]
%%%%%%%%%%%%%%%%

 From the definition of the Schwarzschild radius it follows that
\be
\frac{d r_\sg(u)}{du}=\frac{\pad C}{\pad u}\Big|_{r=r_\sg(u)}+\frac{\pad C(u,r)}{\pad r}\Big|_{r=r_\sg(u)}\frac{d r_\sg(u)}{du},
\ee
hence the condition (i) implies
\be
\big(1-w(u,r_\sg)\big)\frac{dr_\sg}{du}<0,
\ee
sharpening the bound on $w$ to $w(u,r_\sg)<1$ and ensuring  that $dr_\sg/du<0$.

The shell's trajectory is given by $U_\pm(\tau)$, $R_\pm(\tau)$. The first junction condition again identifies $R_+\equiv R_-=:R$, and the
{$v_\mu v^\mu=-1$} constraint  now gives
\be
\bar E ^2F\dot U^2+2\bar E\dot U\dot R=1,
\ee
where $F:=f(U,R)$ and $\bar E:=\exp\big(h(U,R)\big)$. We will not use the second junction condition and equations of motion. Hence we do not assume anything about shell's composition beyond it being massive.

The counterpart of Eq.~\eqref{udot} is
\be
\dot U=\frac{- \dot R+\sqrt{ \dot R^2+F}}{\bar E F} \label{udotg}
\ee
and continuity of $h$ (assumption (ii)) ensures that for finite $U$ the function $\bar E$ is finite.

From Eq.~\eqref{udotg} it follows that  {$\dot U>2|\dot R|/\bar E F$, } hence for $w(u,r_\sg)<1$
\be
\dot U>  \frac{2r_\sg}{\bar E(1-W)}\frac{|\dot R|}{x},
\ee
 where $W(U):=w(U,r_\sg(U))$.  As a result the rate of approach to the Schwarzschild radius for small $x$ is bounded by
 \be
 \dot{x}=\dot R-\dot r_\sg\big(U(\tau)\big)>\dot R(1-\epsilon_*/x), \label{xdotgen}
 \ee
 where
 \be
 \epsilon_*=\frac{2}{\bar E}\frac{r_\sg}{1-W}\left|\frac{d r_\sg}{d U}\right|
 \ee
sets the horizon avoidance scale.

 Putting all these consideration together we conclude that
 \be
 \epsilon_*(\tau)>0,% \qquad \tilde\epsilon_*>0
 \ee
 provided the evaporation is ongoing. Consequently the same alternative  results,  and the Schwarzschild radius is not reached in a finite proper time. 
 
 Assumption (i) is essential for this result: if the evaporation halts at some finite time, $\pad C(u,r_\sg)/\pad u=0$  while $C(r,u)>0$  for $u>\tilde u$, effectively restoring classical gravitational collapse of the shell,  then it will cross the  diminished Schwarzschild radius $\tilde r_\sg=r_\sg(\tilde u)$ in finite proper time.

We can consider the same problem using Schwarzschild coordinates $(t,r)$. The geometry outside the shell is described by the metric \cite{visser:14}
\be\label{sch-co}
ds^2=-k(t,r)^2f(t,r)dt^2+f(t,r)^{-1}dr^2+r^2d\Omega
\ee
where $k(t,r)$ is some function and
\be
f(t,r)=1-C(t,r)/r.
\ee
 Similarly to the previous case we assume  that (i) $ 0\le C < \infty$, with $C>0 $ for $0< t < t_E$, and $\partial C/ \partial t < 0$ as long  as $C>0$; (ii) $k(t,r)$ is continuous; (iii) the metric has one coordinate singularity, an infinite red-shift surface at $f(t,r) = 0$.  This surface is located at the Schwarzschild radius $r_\sg(t)$, given implicitly
by $r_\sg\equiv C(t, r_g)$.
The mass function is now given by
\be
C(t,r)=r_\sg(t)+w(t,r)\big(r-r_\sg(t)\big), \label{sch-co-C}
\ee
and the shell's trajectory is parameterized as $\big(T(\tau), R(\tau)\big)$. The velocity components satisfy
\be
\dot{T}=\frac{\sqrt{F+\dot R^2}}{KF},
\ee
where $K=k\big(T(\tau),R(\tau)\big)$. The Schwarzschild radius is avoided analogously to the previous case, according to $1/x$ with the   scale
\be
\epsilon_*=\frac{1}{K}\frac{r_\sg}{1-W}\left|\frac{d r_\sg}{d T}\right|, \label{epstasch} %\qquad \tilde\epsilon_*^2=\frac{ r_\sg}{\bar E|W_r|}\left|\frac{d r_\sg}{d T}\right|,
\ee
with $W(T):=w\big(T,r_\sg(T)\big)$.

\textit{Spherically-symmetric collapse of   dust.} --- We now consider how emission of the pre-Hawking radiation modifies   Oppenheimer-Snyder collapse \cite{os:39,poisson, ast-bh}. The classical model describes the gravitational collapse of a uniform spherical distribution of dust.   The geometry outside is given by the Schwarzschild metric of Eq.~\eqref{sch-co}. The metric inside the dust is that  of Friedmann-Lema\^{i}tre-Robertson-Walker closed universe,
\be
ds_-^2=-d\tau^2+a^2(\tau)(d\chi^2+\sin^2\chi d\Omega), \label{frw}
\ee
where $\tau$ is the proper co-moving time $\tau$ and $\chi$ the co-moving radial coordinate. The metric parameters are matched via the junction conditions.   The metric \eqref{frw} can be transformed into the $(t,r)$ coordinates of Eq.~\eqref{sch-co}, by, e.g.,  first transforming  to the Painlev\'{e}e-Gullstrand coordinates \cite{blau}. We label layers  by (the initial co-moving)  coordinate $\chi$, so their motion is parameterized as $\big(T_\chi(\tau), R_\chi(\tau)\big)$.

 Even in a classical setting  the shells may cross if the dust is charged  \cite{ori:91}. We assume that (iv) there is no shell crossing when the quantum backreaction is taken into account, so the label $\chi$ is unambiguous.

The simplest scenario involves a pressureless neutral dust that also does not interact  with the radiation field. Here the difference with the standard Oppenheimer-Snyder collapse is in the form of the metric.  {Each dust particle $(T_\chi(\tau),R_\chi(\tau))$ still moves along  a radial geodesic in the geometry   given} by Eq.~\eqref{sch-co}, but with the appropriately modified parameters,
\be
ds_-^2=-k_\chi(t,r)^2f_\chi(t,r)dt^2+f_\chi(t,r)^{-1}dr^2+r^2d\Omega. \label{metchi}
\ee
 The Schwarzschild radius that the layer $\chi$ should cross is $r_\sg^\chi\big(T_\chi(\tau)\big)\equiv C_\chi\big(T_\chi(\tau), r_\sg^\chi\big(T_\chi(\tau)\big)\big)$. Since no specific interior geometry (flat or curved) and no specific form of acceleration $\ddot R(\tau)$ were assumed in deriving  Eqs.~\eqref{xdotgen} and \eqref{epstasch}, there will be no crossing of the Schwarzschild radius, $R_\chi(\tau)=r_\sg^\chi\big(T(\tau)\big)$ for any finite proper time $\tau$.

A more realistic model still involves pressureless dust whose interaction with the radiation field is  summarized by the radiation pressure that is extracted from the stress-energy tensor that is consistent with the metric. Its component $T^r_r$ is  the right hand side of the Einstein equation where the curvature is obtained from the metric \eqref{metchi}. Following discussions that are summarized, e.g. in \cite{wald:01}, the radiation pressure slows collapse down, so there will be still no  crossing.

%\tcg{\sc [The arguments seem to be tight. I explicitly stressed the assumption that shells do not cross.] }

\textit{Conclusions.}---
 Assuming the existence of a pre-Hawking radiation in spherically-symmetric collapse that satisfies (via the corresponding metric) only mild restrictions,  we showed that crossing of the Schwarzschild radius does not take place --- and thus
event horizons do  not form ---   for a generic model of massive thin shell collapse.  This result   extends to  two models of collapsing pressureless dust. An apparent horizon definitely does not appear for thin shells, but is not excluded for the collapse of dust balls (similar to \cite{mersini:14}).

Using the late time form of the evaporation law \cite{fn:98,bmps:95,modern,page:76} we see that for macroscopic super-compact black hole-like objects the typical scale of the horizon avoidance $\epsilon_*\propto 1/r_\sg$ is trans-Planckian.
It provides another indication that the paradoxical aspects of the black hole information problem originate in combining a sharp classical geometry with quantum fields \cite{brust:14, bht:17}.  It also ensures that predictions of this model are observationally indistinguishable from  pure classical collapse \cite{ast-bh,af:13}, at least until the final moments of complete evaporation.

Our results are consistent with the existence of super-compact objects (``black holes without horizon"), such as proposed in \cite{mersini:14}, or fuzzballs \cite{mathur:05}. It is reasonable to conjecture that horizon formation does not take place for collapsing objects in
any configuration, provided that the pre-Hawking radiation is taken into account and  appropriate energy conditions are assumed.

\textit{Acknowledgements.}---
The work of RBM was supported in part by the Natural Sciences and Engineering Research Council of Canada.

%\textit{Acknowledgments.}     VB is supported by the Macquarie Research Fellowship scheme, RBM was supported by the Natural Sciences and Engineering Research Council of Canada.

\end{document}